\documentclass[aps,twocolumn,pre,showpacs,color]{revtex4}
\usepackage{amsmath}
\usepackage{color}
\usepackage{amsfonts}
\usepackage{times}
\usepackage{graphicx}
\begin{document}

\title{On the structure of blue phase III}


\author{O. Henrich$^1$,  K. Stratford$^2$, M. E. Cates$^1$, D. Marenduzzo$^1$}

\affiliation{$^1$ SUPA, School of Physics and Astronomy, University of Edinburgh, Mayfield Road, 
Edinburgh EH9 3JZ, UK \\
$^2$ EPCC, The University of Edinburgh, Mayfield Road,
Edinburgh EH9 3JZ, UK}

\begin{abstract}
We report large scale simulations of the blue phases of cholesteric liquid crystals. Our results suggest a structure for blue phase III, the blue fog, which has been the subject of a long debate in liquid crystal physics. We propose that blue phase III is an amorphous network of disclination lines, {which is thermodynamically and kinetically stabilised over crystalline blue phases at intermediate chiralities}. This {amorphous} network {becomes ordered under an applied electric field, as seen in experiments}.
\end{abstract}

\maketitle


Liquid crystals (LCs) offer prime examples of systems with spontaneously broken symmetry that support topological defects of various types~\cite{Wright1989}. Alongside their technological utility in various optical devices, LCs offer testing grounds for fundamental theories whose counterparts in other fields (such as cosmology, particle physics, and more recently exotic superconductivity~\cite{Other}) are less directly testable. While many aspects of LC physics are by now understood, some, such as the character of the so-called `blue fog' (blue phase III) have remained unresolved despite efforts spanning several decades. 

The simplest liquid crystalline phase is the nematic, in which molecules have a preferred orientational axis (the director) but no translational order. Introduction of molecular chirality causes the director to precess in space. In the simplest resulting phase (the cholesteric) it does so about a single helical axis, thereby creating a 1-D periodic structure whose wavelength is the helical pitch. Locally, however, the ordering remains nematic and the cholesteric therefore supports topological line defects, known as disclinations, in which the director executes a rotation of $\pi$ under a full $2\pi$ rotation around the defect line. Within the cholesteric phase itself, these excitations are absent from the ground state.

In many chiral liquid crystals the transition between the cholesteric and isotropic phases occurs through a cascade of weakly first-order transitions to intermediate structures, known as `blue phases' (BPs) and consisting of self-assembled disclination networks~\cite{Wright1989,Crooker1989}. At high enough chirality, a simple helical structure is less stable locally than a so-called double-twist cylinder (DTC), in which the director field rotates simultaneously in two directions perpendicular to a cylinder axis. However DTCs cannot be smoothly patched together to fill the whole of space; disclination lines are required at the interstices between the cylinders. If an external field is used to align the DTCs, the disclinations can form a simple linear array \cite{PRE}, but more generally they meet at junctions forming a multiply connected network. The resulting structure is highly colored (`blue') since the spacing between defects lies in the optical range. BPs are stable when the free energy gained by creating DTCs instead of a simple helix is enough to compensate the free energy loss due to defect formation. They therefore occur at high chirality (where DTCs are favored) and for $T$ close to but below $T_{IC}$ (the  transition temperature to the isotropic phase) where defect energies are low.
{Long viewed by many as little more than a curiosity~\cite{Wright1989}, BPs have recently emerged as promising materials for photonics and display devices following their stabilisation over a much larger temperature range than previously thought possible (about 60 K compared to $<$1 K)~\cite{stabilised_bp}.} This has revived interest in these fascinating hierarchical materials, in which molecular-scale interactions stabilize topological defects that self-organize into a micron-scale superstructure. 

Three BPs have been observed experimentally (at zero electric field). Two, BPI and BPII, are highly ordered: their disclination networks form a regular cubic lattice, and their physics is well understood. (Another ordered phase, $O_5$, is predicted by theory at high chirality, but not found experimentally.) The last one is known as BPIII, and its structure is one of the unsolved puzzles of liquid crystal physics. Theorists have proposed that BPIII may either be a quasicrystal~\cite{quasicrystal_bp3}, a spaghetti-like tangle of double-twist cylinders~\cite{spaghetti_bp3}, an amorphous state formed by BPII domains~\cite{bp2domain_bp3}, or a metastable phase~\cite{metastable_bp3}. Experiments on the thermodynamics, scattering and electric field response~\cite{expt_bp3,Kitzerow1991,Yang1988} remain inconclusive, although electron micrographs appear to favour an amorphous {structure} and to rule out the quasicrystal model~\cite{Crooker1989}. 

In this work we show that computer simulations can help settle this important physical question. By means of large scale {(supra-unit-cell)} simulations, which enable very accurate comparison of free energies for both ordered and amorphous structures, we provide strong evidence that BPIII is indeed an amorphous network of disclinations.  Remarkably we show that, within a certain window of chirality and with a standard choice of free energy functional detailed below, {\em individual aperiodic structures exist that are more stable than either BPI, BPII, or $O_5$}. This narrow window lies within the observed range of BPIII, and can only become wider, in line with experiments, if one allows for configurational entropy. (Such entropy arises when there exist multiple aperiodic structures of the same free energy, as seems likely here.) Furthermore, we show that an applied electric field orders our aperiodic BPIII candidate into a different, much more ordered blue phase. This concurs with longstanding experiments~\cite{Kitzerow1991,Yang1988} which showed evidence of a field-induced transition to a new phase, BPE, whose structure was not previously identified. 


The local order in the BPs can be described by a traceless, symmetric, second rank tensor, $\mathbf{Q}$ whose equilibrium thermodynamics is governed by the Landau--de Gennes free energy functional ${\cal F}$, whose form, within the one-elastic constant approximation (discussed later) is standard, and specified at \cite{SUP}. Within that approximation, the phase behavior of BPs depends on just three dimensioness parameters, a reduced temperature $\tau$, a reduced chirality $\kappa$ and a reduced field strength ${\cal E}$~\cite{equilibrium}. Expressions  for these in terms of $K$ (the elastic constant), $q_0$ (the cholesteric wave-vector) and other parameters in ${\cal F}$ are given at \cite{SUP}. Good agreement between theoretical~\cite{parameters1} and experimental~\cite{parameters2} phase boundaries is obtained by taking $\tau \propto (T-T_{IC})$ and $\kappa$ linear in the mole-fraction of a chiral component (with respective proportionality constants $\simeq 2$ K$^{-1}$ and $\simeq 2$ for one specific mixture~\cite{parameters1}).

We employ a 3D {hybrid lattice Boltzmann (LB)} algorithm~\cite{parameters1,LBLC,softmatter,PNAS} to solve the 
Beris-Edwards equations. The evolution of the $\mathbf{Q}$
tensor~\cite{beris} is
\begin{equation}
D_t \mathbf{Q} 
= \Gamma  \Bigl( \tfrac{-\delta {\cal F}}{\delta \mathbf{Q}} + \tfrac{1}{3}\, 
\text{Tr} \Bigl( \tfrac{\delta {\cal F}}{\delta \mathbf{Q}} \Bigr) \mathbf{I} \Bigr)  .
\label{eqQevol}
\end{equation} 
Here, $\Gamma$ is a collective rotational diffusion constant and $D_t$ is a 
material derivative for rod-like molecules~\cite{beris}. The term in brackets 
is the molecular field, $\mathbf{H}$, which ensures that $\mathbf{Q}$ evolves 
towards a minimum of the free energy. The fluid velocity field obeys 
the continuity equation and a Navier-Stokes equation with a stress tensor 
generalised to describe liquid crystal hydrodynamics, and discussed 
elsewhere~\cite{LBLC}. 

Though these equations represent the true dynamics, we use them here simply to find free energy minima. Thus, as in previous work  we additionally allow a so-called `redshift' in which the parameters in ${\cal F}$ are dynamically updated at fixed $\tau,\kappa,{\cal E}$~\cite{SUP}. This exploits a scaling among those parameters to ensure that the system is not frustrated by periodic boundaries: in particular, for any cubic BP, a lattice parameter emerges that truly minimizes ${\cal F}$. The accuracies of their computed free energies $F \equiv \min({\cal F})$ are (at least for ${\cal E} = 0$) thus limited only by discretization. This is chosen to fully resolve the defects~\cite{SUP}, whose core energy is finite, and set by ${\cal F}$ itself.

In computing the free energy of periodic structures (BPI, BPII and $O_5$) we apply a perturbation of the appropriate symmetry to a uniform state and then evolve dynamically \cite{SUP}. This delivers an accurate $F$ value in each such phase, of which the lowest can be chosen; but (in common with all other methods for computing free energies of ordered phases) we cannot rule out others of still lower ${\cal F}$. To address BPIII, we also need to generate aperiodic candidates. Here our strategy is similar: we start from various different aperiodic initial conditions, evolve each dynamically, and choose that of lowest ${\cal F}$~\cite{SUP}. As shown below, this beats all three periodic structures within a certain parameter window. Because we cannot exhaust all possible initial conditions, our free energy is an upper bound on aperiodic states; further exploration can thus only widen that stability window. 
In practice, all the aperiodic candidates we generated look similar: fuller exploration is thus unlikely to change our conclusions about the character of BPIII.

To minimize finite size effects, we simulate very large systems (in contrast to ~\cite{parameters1}). Typically we used 128$^3$ lattices, which accomodates 8 half pitches in each direction~\cite{supercomputers}. Selected simulations with 256$^3$ lattices confirm the results which we report below. 
As shown in~\cite{PNAS}, it is easy to generate aperiodic structures by placing a localized nucleus of BPI or II in a cholesteric or isotropic matrix. However, the lowest ${\cal F}$ values we have so far found are instead achieved by initialising the system in the cholesteric phase in the presence of a low density (typically about 1-2 \% in volume) of randomly placed doubly twisted droplets. 
Once initialized, the system is relaxed dynamically until it reaches a quiescent end-state.

For low chirality ($\kappa<1.5$), the initial defects are washed out to leave a cholesteric phase. However, for a large regime of intermediate chiralities ($1.75\le \kappa\le 3$), our simulations show an intriguing dynamics, through which the dilute doubly twisted regions grow and rearrange dramatically to form a whole network of disclinations, which very slowly creeps to an amorphous end-state (Fig.~1, top). Its amorphous character is confirmed by the structure factor $C({\bf k})$ (Fig.~1, bottom). The ring in $C({\bf k})$ is set by the average distance between the branch-points in the defect network, whereas the small residual peaks (which break spherical symmetry) are likely due to residual finite size or periodic boundary effects. The amorphous network approaches equilibrium through very slow local rearrangements of the disclination junctions; the end-state in Fig.~1 is very close to kinetic arrest, in at least a local minimum of ${\cal F}$. 
Notably, four disclination lines meet at most junctions, so that, as suggested by experiment~\cite{expt_bp3}, the structure is locally closer to that of BPII than BPI. 

Remarkably, for larger values of the chirality ($\kappa>3$) our simulations attain a much more regular state, closely resembling $O_5$ which (see below) minimizes our chosen ${\cal F}$ at  very high $\kappa$. (In $O_5$ itself, whose free energy we have computed precisely, eight disclination lines merge at each junction~\cite{SUP}.) Therefore, our methodology is capable in principle of finding a periodic disclination lattice, if kinetically accessible. Given that our equations fully incorporate liquid crystal hydrodynamics, we believe that the kinetic propensity to form a disordered disclination network for intermediate chiralities indeed reflects a real physical property, although our limited simulation times ($\simeq 1$ms~\cite{SUP}) may well exaggerate the stability window of amorphous structures in the $(\kappa,\tau)$ plane. 

\begin{figure}[htp]
\includegraphics[width=0.33\textwidth]{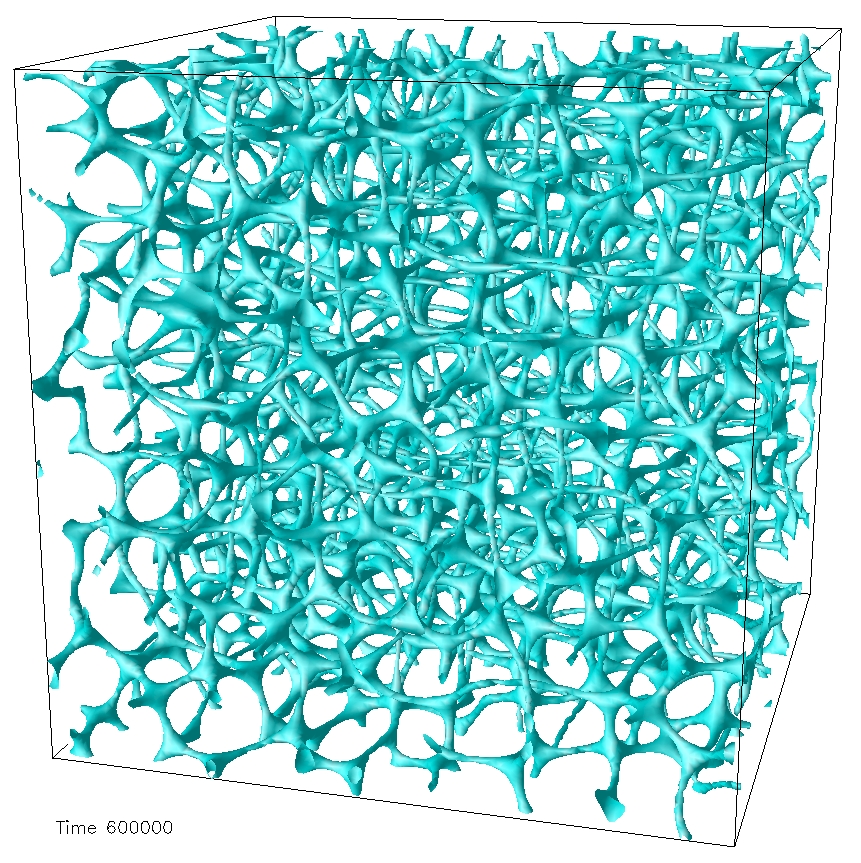}\\
\includegraphics[width=0.2\textwidth]{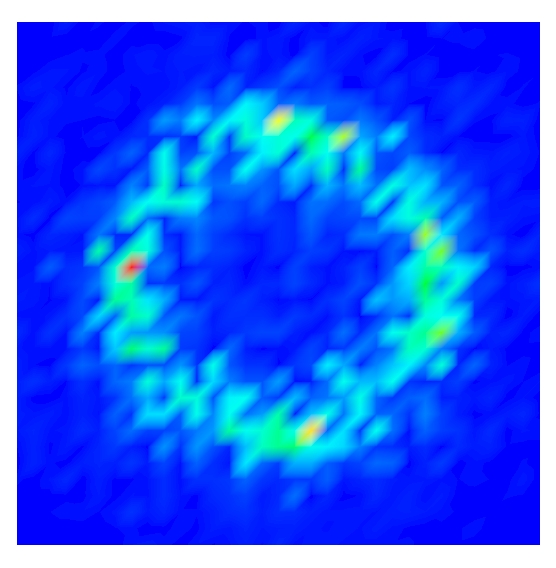}
\includegraphics[width=0.2\textwidth]{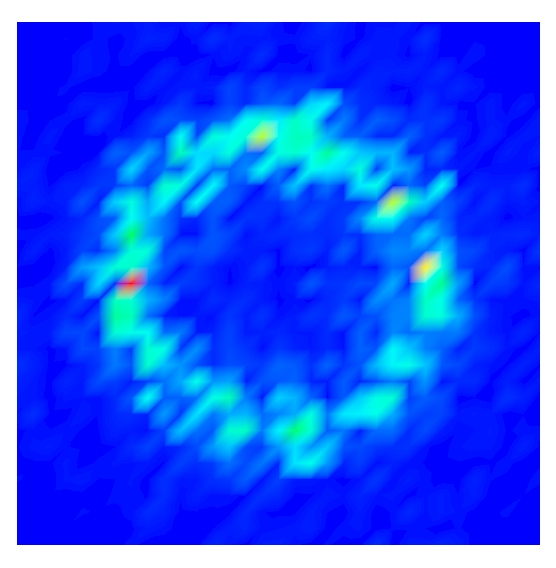}
\caption{Top: End-state disclination network at $\tau=-0.25, \kappa=2.5$: The picture shows the isosurface $q({\bf r})=0.12$, with $q$ the largest eigenvalue of ${\bf Q}$. Inside each tube is a disclination line (on which $q$ takes a minimum value).
Bottom: Structure factor $C(\vec{k}) \equiv |q({\bf k})|^2$, on cuts along $k_y=0$ (left) and $k_x=0$ (right) with wavevectors $k_x/q_0,k_z/q_0\in [-4,4]$ and $k_y/q_0,k_z/q_0\in [-4,4]$ respectively.}
\label{bp3-equ}
\end{figure}

A crucial question is whether our BPIII candidate structure is only kinetically, or also thermodynamically stable. To answer this, we have carefully compared $F(\kappa,\tau)$ for BPIII with those of BPII and $O_5$, for a range of chiralities $\kappa$ at selected values of $\tau$. (BPI is not competitive in the $\kappa$ range of interest here.) As an example, curves for $F(\kappa,-0.25)$ are shown in Fig.~2. We see that there is a small but finite chirality window in which the BPIII-network is the thermodynamic equilibrium phase. We find it remarkable that any single aperiodic structure can outcompete such periodic ones when minimizing the relatively simple Landau-de Gennes free energy. In particular, this minimization takes no account of order-parameter fluctuations about the local minimum, which might help stabilize BPIII~\cite{stark}, nor the configurational entropy associated with having many such minima.
(These neglected contributions should be small -- of order $k_BT$ per unit cell while the free energy differences in Fig.~2 are typically of order 100-1000 times larger~\cite{SUP}.) However, the thermodynamic stability window in BPIII is smaller than the one we find kinetically  -- in agreement with the expectation that amorphous structures should form more easily dynamically than highly ordered counterparts, which require long and complicated process to annihilate any dislocations and overcome relatively high energy barriers. The window of thermodynamic stability might also be broadened by relaxing the one elastic constant approximation, particularly if this raises the free energy of $O_5$ relative to the other states. This certainly merits further study, given the experimental absence of that phase.

\begin{figure}[htp]
\includegraphics[width=0.4\textwidth]{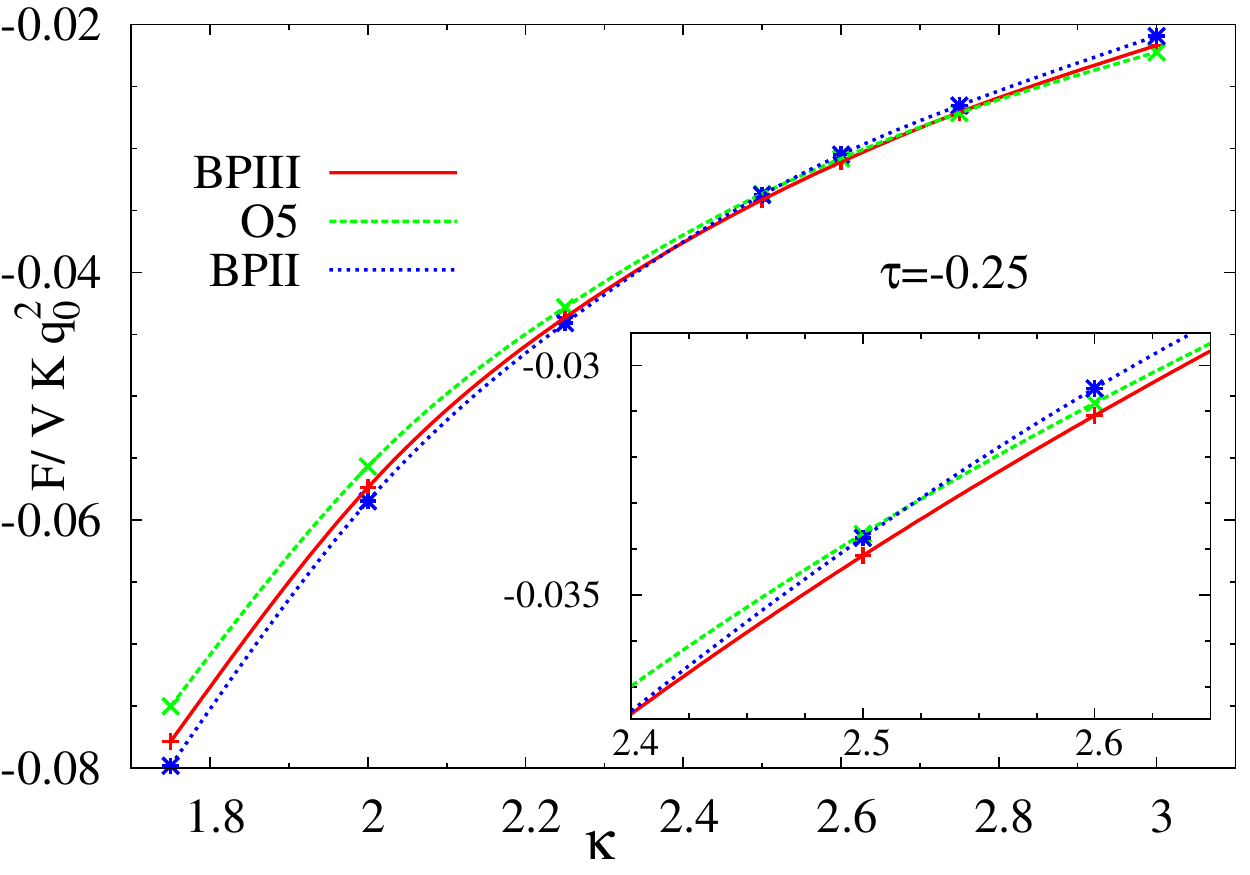}
\caption{Free energy densities ($V$ is the volume) vs. chirality $\kappa$ at $\tau=-0.25$ for BPII, our candidate for BPIII, and $O_5$. The inset shows a blow-up of the region where our BPIII is the equilibrium phase. Our free energy densities are expressed in units of $Kq_0^2$ (see~\cite{SUP} for the mapping between simulation and real units).}
\label{fe}
\end{figure}

All the above arguments support our assignment of the amorphous network seen in Fig.~1 as the theoretically elusive BPIII, stable at larger $\kappa$ than BPII, but locally similar to it~\cite{Crooker1989}. 
Experiments on BPIII in electric fields also point to a field-induced transition between BPIII and another structure which has been named BPE~\cite{Yang1988,Kitzerow1991}. BPE was found to give rise to a sharper peak in the scattering data consistent with an enhanced ordering {(unfortunately Refs.~\cite{Yang1988,Kitzerow1991} do not provide further experimental data on the structure of BPE).}
{It is therefore interesting to ask what happens when our amorphous structure is subjected to an electric field.} In our simulations we can follow the evolution of the {\bf Q} tensor and of the disclination network in an external field ${\cal E}$, and also compute $C({\bf k})$; if a crystalline phase emerges, it will exhibit Bragg peaks.

By stepwise increasing ${\cal E}$ from 0 to 0.65, we found that the disclination network in Fig.~1 melts away leaving a nematic state. (This holds for positive dielectric anisotropy~\cite{SUP}.) We then performed a long simulation at ${\cal E}=0.55$, close to but below the threshold beyond which the system becomes nematic. The time evolution of our BPIII, stable at zero field, is shown in Fig.~3. This shows an interesting rearrangement of the defect texture, through which the junction points rotate and finally reconstruct to yield a topologically distinct phase, with helical disclinations lying along layers stacked perpendicular to the electric field (which is vertical). The disclination lines in two consecutive layers are turned by 90 degrees, so that they show a square arrangement when viewed along the field direction. Within a layer two adjacent disclination lines are staggered. Again in agreement with experiments, then, we find that our {amorphous BPIII candidate} undergoes a field-induced transition to an ordered disclination network which we provisionally identify as BPE. {It would be interesting to perform additional experiments to test such an identification. Our candidate BPE appears} related to, but distinct from, other field- or confinement-induced BPs previously reported~\cite{PRE,confine}, with strong crystalline order (albeit with some defects remaining) as confirmed by its structure factor.

\begin{figure}
\includegraphics[width=0.22\textwidth]{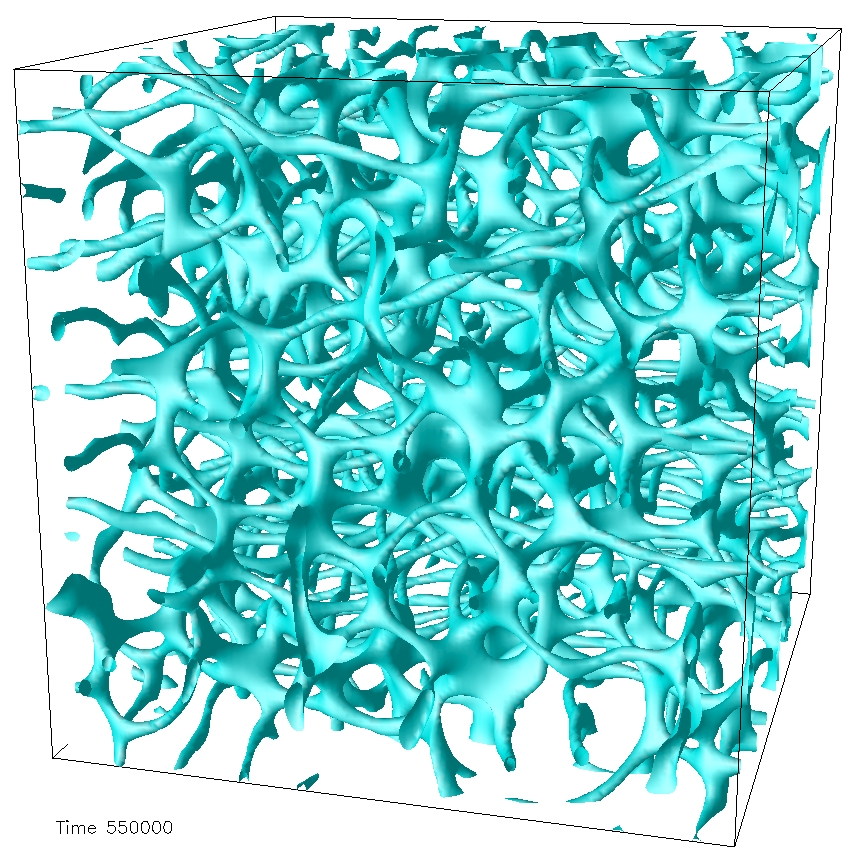}
\includegraphics[width=0.22\textwidth]{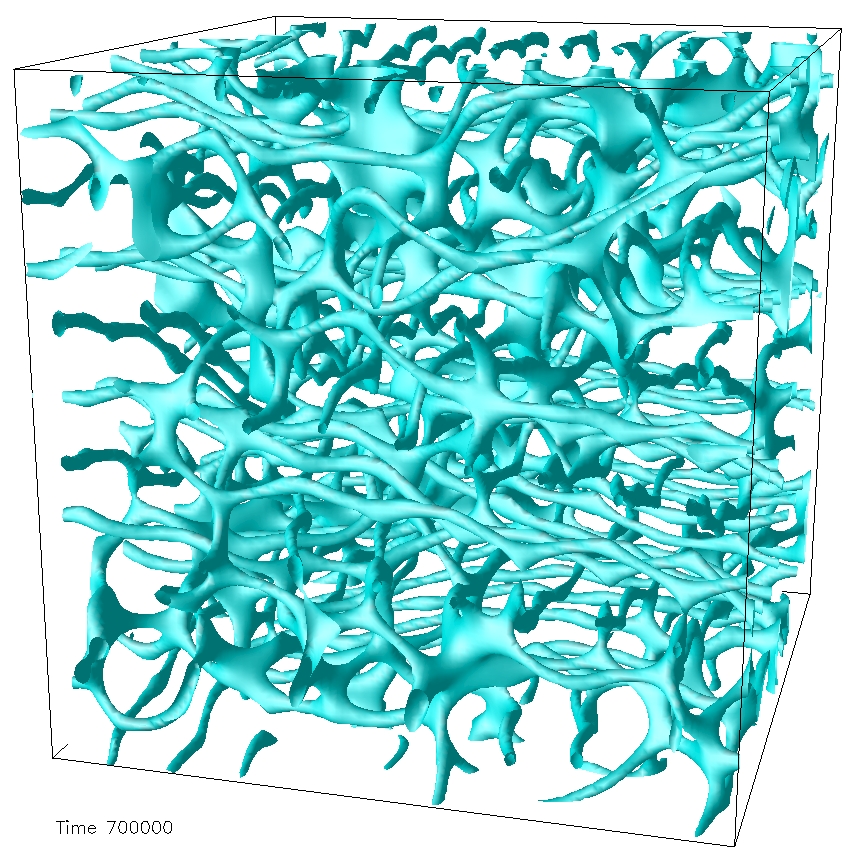}\\
\includegraphics[width=0.22\textwidth]{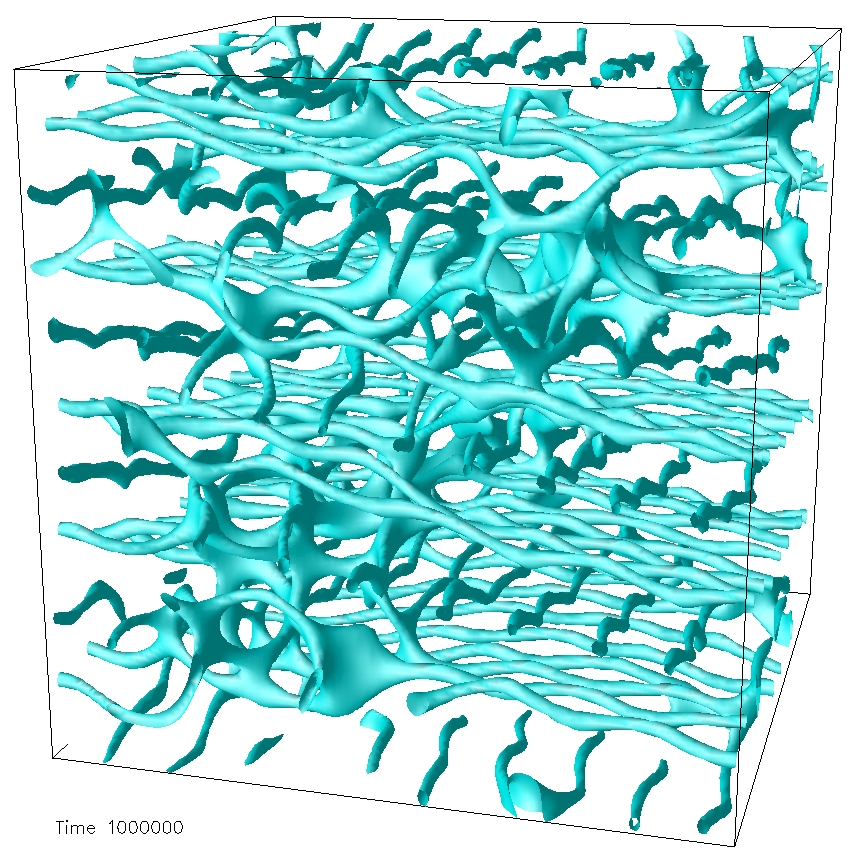}
\includegraphics[width=0.22\textwidth]{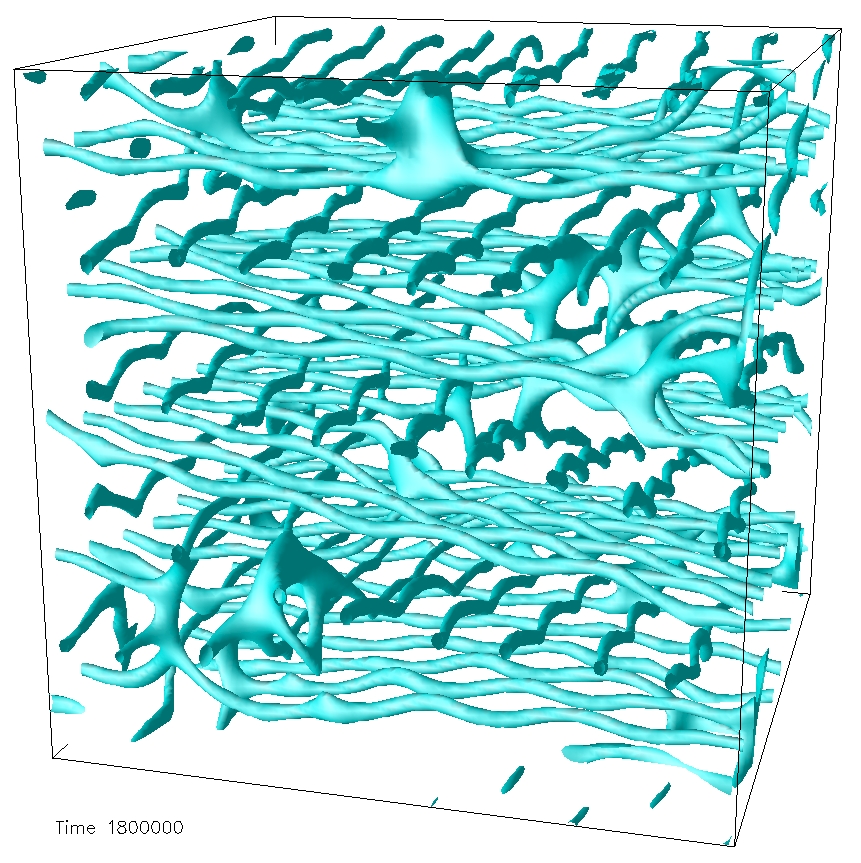}\\
\includegraphics[width=0.2\textwidth]{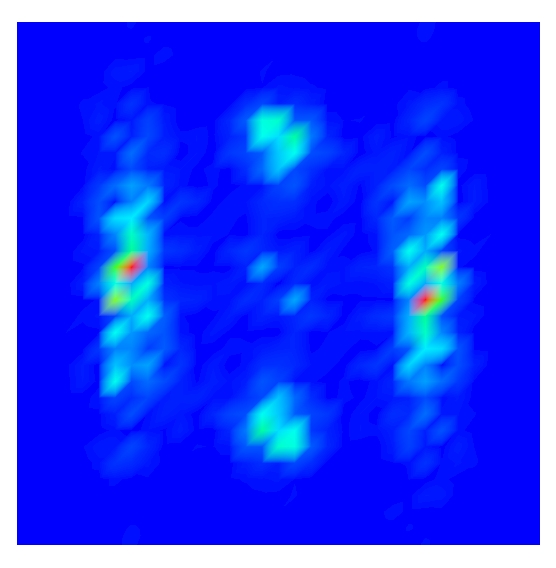}
\includegraphics[width=0.2\textwidth]{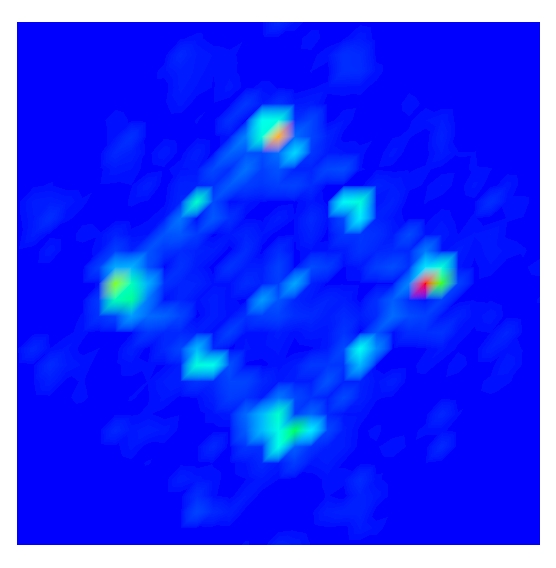}
\caption{Isosurfaces of $q=0.12$ at $\tau=0.2, \kappa=2.0, {\cal E}=0.55$: The sequence shows the transition from the amorphous BPIII network to a field-induced BPE. The field direction, $z$, is vertical. 
Times shown are $t \simeq 0.15, 0.3, 0.6$ and $1.4$ ms (taking one timestep $\simeq 1$ ns).
Bottow row: Structure factor $C(\vec{k})$ at the end of the run. The pictures show a cut through the 3D data at $k_y=0$ (in the $(k_x,k_z)$ plane, left) and $k_z=0$ (in the $(k_x,k_y)$ plane, right), with scales as in Fig. 1.}
\label{bp3-e0o55}
\end{figure}


In conclusion, we have presented large scale simulations of the cholesteric blue phases. We have given strong evidence that in a region of the temperature--chirality plane, instead of a periodic BP, an aperiodic one is selected kinetically. 
Moreover, and strikingly, within the (one-elastic constant) Landau - de Gennes free energy~\cite{SUP}, we have found a finite window of chirality in which this amorphous network is thermodynamically more stable than the competing crystalline blue phases, BPI, BPII and $O_5$. These facts suggest that {our kinetically and thermodynamically stable network is none other than the blue fog, BPIII.} This view is strengthened by the study of the field response of this structure, which reconstructs into an ordered phase at intermediate values of the electric field, slightly smaller than those at which the disclination network melts to give a {field-oriented} nematic phase. 

We believe our simulations {shed important light on the} structure of the blue fog, which we propose is an amorphous disclination network which is locally close to BPII: this view is perhaps intermediate between the ``BPII domain'' \cite{bp2domain_bp3} and the ``double twist tangle'' \cite{spaghetti_bp3} models which were proposed in the early 1980's, and is at odds with some of the subsequent structures proposed for this phase. Several open questions remain or are stimulated by our results. Firstly, {one should} characterise statistically the phase transition from the blue phase to the isotropic phase: this may {be possible by adding thermal noise to simulations like the ones reported here}. Secondly, use of more accurate scattering and visualisation techniques should allow tests of our candidate structure for BPE (see Fig.~3). Finally, if the large energy scales that separate the various BP topologies (of order $100-1000k_BT$ per unit cell \cite{SUP}) also control reconstruction of the defect network within BPIII itself, one may expect that in some materials, even when thermal noise is allowed for, such reconstruction cannot be achieved on any reasonable timescale. If so, BPIII will represent another elusive entity: an amorphous state of globally minimum free energy that is kinetically arrested --- or an `equilibrium glass'. 

We thank A. N. Morozov for useful discussions. 
This work was supported by EPSRC {grants EP/E045316/1 and EP/E030173/1. MEC is funded by the Royal 
Society.}

\appendix

Here we provide further information on the free energy functional and the control parameters $\tau, \kappa, {\cal E}$ (section A); on resolution and  on `redshift', i.e., the parameter-dependent unit cell size of blue phases (section B); on the initial conditions chosen for runs presented in the main text (sections C and D); on the conversion of simulation parameters to physical units (section E); on the free energy comparison between different blue phases (section F); and on the effect on field-response of changing the sign of the dielectric anisotropy (section G).

\subsection{Free energy functional, chirality and reduced temperature}

The thermodynamics of cholesteric blue phases can be described via a 
Landau-de Gennes free energy functional ${\cal F}[{\bf Q}]=\int d^3{\bf r} f({\bf Q}({\bf r}))$.
The free energy density we use, following \cite{Wright1989}, is:
\begin{eqnarray}
f({\bf Q})&=&\frac{A_0}{2}\left(1-\frac{\gamma}{3}\right)Q^2_{\alpha \beta}-\frac{A_0\gamma}{3}Q_{\alpha \beta} Q_{\beta \gamma}Q_{\gamma \alpha} + \frac{A_0\gamma}{4} (Q^2_{\alpha \beta})^2  \nonumber\\
&+&\frac{K}{2}(\varepsilon_{\alpha \gamma \delta} \partial_\gamma Q_{\delta \beta} + 2 q_0 Q_{\alpha \beta})^2+ \frac{K}{2}(\partial_\beta Q_{\alpha \beta})^2  \nonumber
\\
& - & \tfrac{\varepsilon_a}{12\pi} 
	   E_{\alpha} Q_{\alpha \beta} E_{\beta}
\label{free}
\end{eqnarray}
Here repeated indices are summed over and $Q^2_{\alpha\beta}$ stands for $Q_{\alpha\beta}Q_{\alpha\beta}$ etc.
The first three terms are a bulk free energy density whose overall scale is set by $A_0$ (discussed further below); $\gamma$ is a control parameter, related to reduced temperature. Varying the latter in the absence of chiral terms ($q_0=0$) gives an isotropic-nematic transition at $\gamma = 2.7$ with a mean-field spinodal instability at $\gamma = 3$.
The next two terms in Eq.\ref{free} describe distortions of the order parameter field. As is conventional \cite{Wright1989,deGennes} we have assumed that splay, bend and twist deformations of the director are equally costly; $K$ is then the one elastic constant that remains. The parameter $q_0$ is 
related via $q_0=2\pi/p_0$ to the pitch length, $p_0$, describing one full turn of the director in the cholesteric phase.

The remaining term accounts for the coupling to the electric field, $E_{\alpha}$, and $\varepsilon_a$ is the dielectric anisotropy. In the text we considered the case $\varepsilon_a>0$, whereas below we show what happens for negative dielectric constant.

The phase behaviour of blue phases (in the absence of noise due to 
thermal fluctuations, i.e., as computed by minimization of ${\cal F}$) may be shown to only depend on the 
following dimensionless parameters~\cite{Wright1989}:
\begin{eqnarray}
\tau&=&{27(1-\gamma/3)}/{\gamma}\label{tau}\\ \nonumber
\kappa&=&\sqrt{{108\ K\, q_0^2}/(A_0\, \gamma)}\label{kappa}\\ \nonumber
{\cal E}^2&=&(27 \varepsilon_a E_\alpha E_\alpha)/(32 \pi A_0 \gamma) \label{e}.
\end{eqnarray}
as used in the main text. These are referred to as reduced temperature, chirality and reduced field 
respectively. Note that if $f$ is made dimensionless, $\tau$ appears as prefactor of the term quadratic in ${\bf Q}$, whereas $\kappa$ quantifies the ratio between bulk and gradient terms.

\subsection{Resolution, unit cell size and `redshift'}

In our simulations, each disclination core spans a few (typically 3-4) lattice sites. To confirm that the defect core is adequately resolved, we have run selected simulations in which the resolution was doubled or quadrupled, and found very similar values for the free energy. On this basis we assess that the typical error due to discretization for our main runs is much smaller than the free energy difference between BPs.

In cubic BPs it is known that the equilibrium unit cell size is slightly different from (and typically slightly larger than) the half pitch of the cholesteric liquid crystal, ${\pi}/{q_0}$. To account for this, a `redshift' factor $r$ is introduced \cite{parameters1} whose variation effectively allows free adjustment of the BP lattice parameter, $\Lambda \to \Lambda/r$, despite the use of periodic boundary conditions in our simulations.  
That is, to avoid changing the size of the simulation box, redshifting is performed by an equivalent rescaling of the pitch parameter and elastic constant, $q_0\to q_0/r$ and $K\to K \,r^2$. In simulations aimed solely at free energy minimization, it is legitimate to make $r$ a dynamic parameter and update it on the fly to achieve this. At each time step one then selects the value of $r$ 
for which the physical free energy density is lowest. More details on this procedure are given in Ref.~\cite{parameters1} where this scheme was first proposed. 

We have used this dynamic updating of the redshift (i) in all single unit cell runs used to map out the free energy curves, and (ii) in multi-unit cell dynamical runs after the final network was formed. Redshifting is essential in order to get accurate estimates for the free energy of the {\it cubic} blue phases, whose relative free energy differences between phases is less than a percent of their typical values. With redshifting, the accuracy attainable is in principle limited only by discretization error; without it, the dominant error is from mismatch between the preferred unit cell size and the periodic boundary conditions.
For amorphous networks, those conditions enforce a periodic structure (on the box scale) with the aperiodic structure within that; redshifting likewise avoids any incommensurability between the two length scales. Because of the constrained periodicity, the computed free energy for amorphous structures will always be overestimated, despite which we find a window of thermodynamic stability for these with free energies lower than the known ordered phases. (As usual we cannot rule out a still lower free energy for some other, unknown, ordered phase.) Our use of redshift on all phases thus means that the true thermodynamic stability window of BPIII can only be {\em wider} than the one found here.

\subsection{Initial conditions: cubic blue phases}

To accurately compute the free energy of cubic blue phases from single unit cell simulations, we must start from an initial condition of the correct symmetry and disclination line topology. To this end, as discussed in Ref.~\cite{equilibrium}a, we can adopt the form for ${\bf Q}(x,y,z)$ corresponding to the limiting form for each phase in the high chirality limit~\cite{Wright1989}.

For BPI, with $q_0'=\sqrt{2} q_0$, we then have initially:
\begin{eqnarray}
Q_{xx} &\simeq&  -2 \cos(q_0' y)\sin(q_0' z)+\sin{(q_0' x)} \cos{(q_0' z)} 
\nonumber \\  &+& \cos{(q_0' x)} \sin{(q_0' y)}\nonumber\\
Q_{xy}&\simeq& \sqrt{2} \cos{(q_0' y)} \cos{(q_0' z)}+\sqrt{2} \sin{(q_0' x)} \sin{(q_0' z)} \nonumber \\ &-& \sin{(q_0' x)} \cos{(q_0' y)}\nonumber\\
Q_{xz} &\simeq& \sqrt{2}\cos(q_0' x) \cos(q_0' y)+\sqrt{2} \sin(q_0' z) \sin(q_0' y) \nonumber \\  &-& \cos(q_0' x) \sin(q_0' z)\nonumber\\
Q_{yy}&\simeq&-2 \sin(q_0' x)\cos(q_0' z)+\sin(q_0' y)\cos(q_0' x) \nonumber \\
&+&  \cos(q_0' y)\sin(q_0' z)\nonumber\\
Q_{yz}&\simeq& \sqrt{2} \cos(q_0' z) \cos(q_0' x)+ \sqrt{2}\sin(q_0' y) \sin(q_0' x) \nonumber \\ &-& \sin(q_0' y)\cos(q_0' z).
\end{eqnarray}
Likewise, for BPII we start from:
\begin{eqnarray}
Q_{xx}&\simeq& \cos(2 q_0 z)-\cos(2 q_0 y)\nonumber\\
Q_{xy}&\simeq& \sin(2 q_0 z)\nonumber\\
Q_{xz}&\simeq& \sin(2 q_0 y)\nonumber\\
Q_{yy}&\simeq& \cos(2 q_0 x)-\cos(2 q_0 z)\nonumber\\
Q_{yz}&\simeq& \sin(2 q_0 x).
\end{eqnarray}
Finally, for $O_5$ we start from:
\begin{eqnarray}
Q_{xx}&\simeq& 2 \cos(\sqrt{2} q_0 y) \cos(\sqrt{2} q_0 z) \nonumber 
\\ & - & \cos(\sqrt{2} q_0 x) \cos(\sqrt{2} q_0 z) \nonumber \\ &-&
\cos(\sqrt{2} q_0 x) \cos(\sqrt{2} q_0 y)\nonumber\\
Q_{xy}&\simeq& \sqrt{2}\cos(\sqrt{2} q_0 y) \sin(\sqrt{2} q_0 z) \nonumber \\
& - & \sqrt{2} \cos(\sqrt{2} q_0 x) \sin(\sqrt{2} q_0 z)\nonumber \\
&-&\sin(\sqrt{2} q_0 x) 
\sin(\sqrt{2} q_0 y)\nonumber\\
Q_{xz}&\simeq& \sqrt{2} \cos(\sqrt{2} q_0 x) \sin(\sqrt{2} q_0 y) \nonumber \\
&-&\sqrt{2} \cos(\sqrt{2} q_0 z)\sin(\sqrt{2} q_0 y)\nonumber \\
&-&\sin(\sqrt{2} q_0 x)\sin(\sqrt{2} q_0 z)\nonumber\\
Q_{yy}&\simeq& 2 \cos(\sqrt{2} q_0 x) \cos(\sqrt{2} q_0 z) \nonumber \\
&-& \cos(\sqrt{2} q_0 y) \cos(\sqrt{2} q_0 x)\nonumber \\
&-& \cos(\sqrt{2} q_0 y) \cos(\sqrt{2} q_0 z))\nonumber\\
Q_{yz}&\simeq& \sqrt{2} \cos(\sqrt{2} q_0 z) \sin(\sqrt{2} q_0 x) \nonumber \\
&-& \sqrt{2} \cos(\sqrt{2} q_0 y) \sin(\sqrt{2} q_0 x)\nonumber \\
&-& \sin(\sqrt{2} q_0 y) \sin(\sqrt{2} q_0 z).
\end{eqnarray}

This procedure was already discussed in Refs.~\cite{parameters1} and \cite{equilibrium}a, where additional details can be found. Fig.~4 shows the unit cell of $O_5$ obtained after a full numerical minimisation (with redshift) starting from the high chirality limit detailed above. To ensure that equilibrium was reached, before ending a run we required that the free energy did not change, in two successive measurements, by more than an accuracy threshold (typically chosen as less than 0.1\%).

\begin{figure*}
\includegraphics[width=0.45\columnwidth]{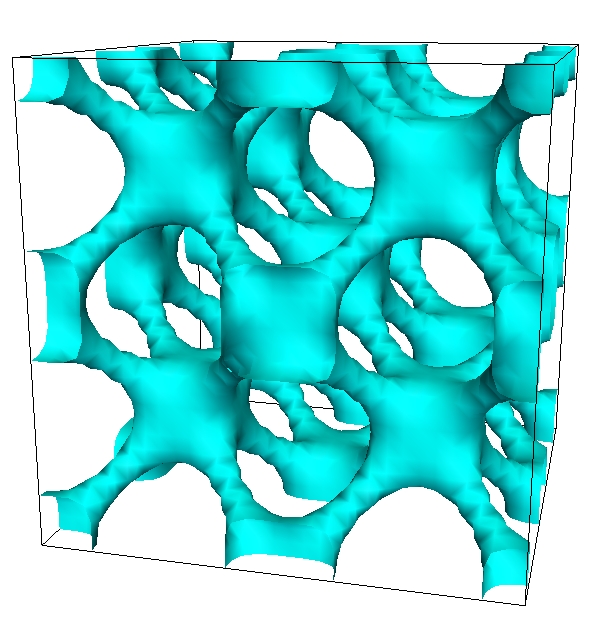}
\caption{Image of the unit cell of $O_5$ after free energy minimization.}
\end{figure*}

In line with previous theoretical and numerical calculations~\cite{Wright1989,equilibrium}, we find that $O_5$ is the equilibrium structure at high enough chirality (see Fig.~2 in the main text). This is therefore expected within the free energy functional formalism that we use. Given that $O_5$ is not observed experimentally, it would be interesting to study its window of thermodynamic stability by relaxing the one elastic constant approximation employed in this work. 

\subsection{Initial conditions: blue phase III}

In the main text we presented configurations for our candidate blue phase III which were obtained by initialising the system in the cholesteric phase in the presence of a low density (typically about 1-2 \% in volume) of randomly placed doubly twisted droplets. One may therefore wonder what happens when different initial conditions are selected.

To address this question, we have performed additional runs starting from a different initial configurations, in which the system was initialised in the {\it isotropic} phase (rather than in the cholesteric phase as in the main text), again in the presence of a low density of randomly placed doubly twisted droplets. Fig.~5 shows the initial and final disclination network in this case -- it can be seen that the final state is very similar to the one shown in the main text. A minor difference is that, in the case of an isotropic background, we found that the initial volume density of doubly twisted impurities needs to be slightly larger for the network to grow. In no case did we find a free energy by this route that was lower than the one resulting from the cholesteric initial condition.
As shown in~\cite{PNAS}, finally, it is also possible to generate locally stable aperiodic structures by placing a localized nucleus of BPI or II in a cholesteric or isotropic matrix. These were found to have higher free energies than those reported here (but comparable with them). 

\begin{figure*}
\includegraphics[width=0.45\columnwidth]{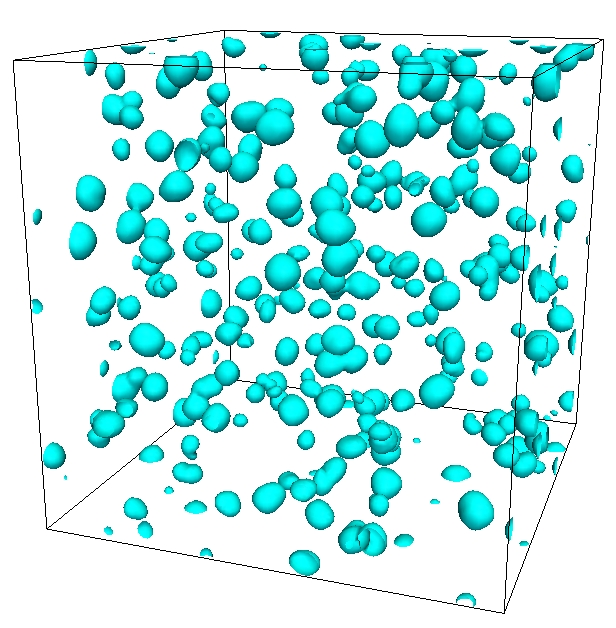}
\includegraphics[width=0.45\columnwidth]{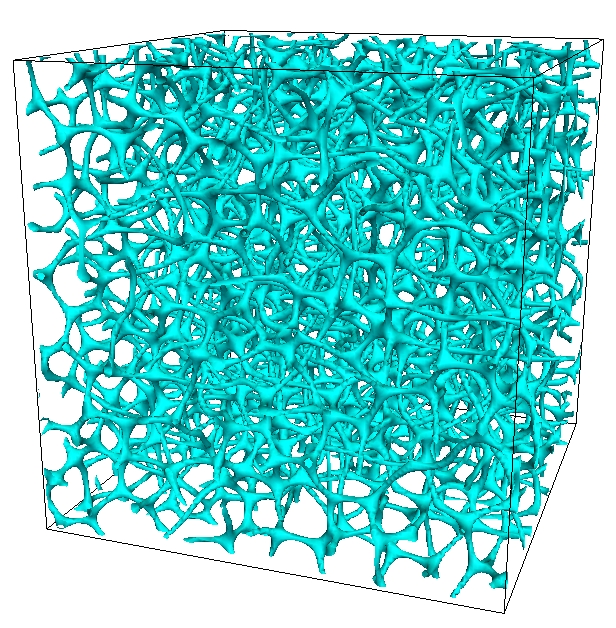}
\caption{Initial (left) and final (right, after 300,000 lattice Boltzmann steps) configuration for a run which is started with randomly placed and oriented doubly twisted droplets, inside an {\it isotropic} background, as opposed to a cholesteric one as presented in the main text. Also in this case we observe a final amorphous network, very similar to the one shown in Fig.~1 of the main text.}
\end{figure*}

To summarise, the kinetic stabilisation of amorphous networks occurs from a wide range of initial conditions, including doubly twisted impurities inside a cholesteric (main text) or isotropic (background), and cubic blue phase nuclei inside cholesteric or isotropic phases. Such a predicted stabilisation is therefore a very robust result of our simulations.  

\subsection{Parameter mapping to physical units}

Here we describe how simulation parameters are related to physical quantities in real BP materials. In order to get from simulation to physical units, we need to calibrate scales of length, energy and time. 
We follow the methodology of~\cite{PNAS,LBLC}.
First we define a set of lattice Boltzmann units (LBU) in which the lattice parameter $\ell$, the time step $\Delta t$, and a reference fluid mass density $\rho_0$ are all set to unity. This is the set of units in which our algorithm is actually written. 
The length scale calibration is straightforward, and fixed by the cholesteric pitch $p_0$, which
is typically in the 100-500 nm range \cite{Wright1989}. More precisely, in our simulations we set the unit cell of BPI/II to be 16 LBU;  this gives good resolution without wasting resource. Therefore the LBU of length (one lattice site) corresponds to, say, 10nm in physical space.

To get an energy scale, we use the measurements cited in Appendix D of \cite{Wright1989} which, as shown in ~\cite{PNAS} suggest:
\begin{equation}
\frac{27}{2 A_0\gamma} \sim 2-5 \times 10^{-6}
{\rm J}^{-1}\, {\rm m}^{3}.
\end{equation}
From this relation (given that $\gamma \simeq 3$ close to the isotropic/cholesteric transition) we obtain that $A_0 \simeq 10^6$ Pa. On the other hand, our simulations typically use a value of $A_0\simeq 0.01$ LBU. This requires that the LBU of free energy density is about $10^8$ Pa in SI units. Therefore our choice of elastic constant, $K = 0.01$ LBU, equates to $K= 10^{-10}$ N, corresponding to a Frank elastic constant $K/2 = 50$ pN which is sensible. 

For dynamical studies one can use the above results to calibrate the lattice Boltzmann time unit, and then crosscheck that the resulting fluid density gives acceptable Reynolds numbers. This is indeed confirmed in \cite{PNAS} for our simulations, which choose an order parameter mobility $\Gamma = 0.3$ LBU in the notation of that paper, and a fluid viscosity  $\eta = 1-2$ LBU. However, such considerations are not crucial in the current context where the dynamics need not be realistic, so long as they accurately minimize ${\cal F}$. To summarize the results above (and the additional discussion in \cite{PNAS}): our simulations faithfully represent experimentally realisable BP-forming materials, subject to the interpretation of the LBU for length, energy density and time are close to 10 nm, 100 MPa and 1 ns respectively.

\subsection{Free energy comparisons}

Table I shows the free energy obtained for our candidate blue phase III, blue phase II and $O_5$, in a temperature--chirality range somewhat larger than the one presented in the main text. (BPI is not competitive throughout this region.)
At two points in parameter space, the free energies (in bold) are lower for BPIII than any other structure. 
For all other parameter sets in the Table, with the exception of one marked ``CHOL'',  the BPIII candidate remains kinetically stable (when initiated from double-twist droplets in a cholesteric phase) even though it has higher free energy than one or more ordered phases. 

A difference in the tabulated values of $\Delta \tilde f =  0.1$ between phases (that is, $\Delta f = 10^{-6}$ LBU) corresponds, with the parameter mapping detailed above, to a free energy difference per unit cell of around $10 k_BT$ per unit cell. This assumes a cell size of 160 nm, which is at the low end for blue phases; for a cell of say 500 nm the difference would be several hundreds $k_BT$ per cell. Adding thermal noise in the simulations (thereby weighting states by the Boltzmann factor $\exp[-{\cal F}/k_BT]$ rather than simply minimizing ${\cal F}$) would be expected to create entropic shifts in the free energies of order $k_BT$ per unit cell, reflecting configurational entropy and/or local order parameter fluctuations.
These shifts would likely favor the amorphous BPIII structure, but,
by these estimates, will do so only marginally.

The above scaling of free energy densities via $k_BT$ per unit cell is useful when discussing fluctuation effects but extraneous at the level of free energy minimization using the Landau - de Gennes functional, in which $k_BT$ does not appear. In that context, it is more natural to express the differences in free energy density between phases in units of either $A_0$ or $Kq_0^2$. Since it relates to more easily measured quantities the latter scaling is preferable, and is followed in Fig.~2 of the main text.

\begin{table*}[h]
\begin{tabular}{|c|c|c|c||c|c||c|c||c|c|}
\hline
$\tau$ & $\kappa$ & $\gamma$ & $A_0\cdot10^3$ & $r$ - BPIII & $\tilde f(r)$ - BPIII & $r$ - O5 & $\tilde f(r)$ - O5&$r$ - BPII& $\tilde f(r)$ - BPII\\
\hline
\hline
0 & 2 &3  &6.9396 &1.097 &-3.104123 &1.009 &-3.047634 &0.908 &-3.161573 \\
0 & 2.5 &3  &4.4413 &1.090 &-1.804606 &1.016 &-1.828106 &0.914 &-1.791576 \\
0 & 2.75 & 3 &3.6705 &1.049 &-1.415262 &1.018 &-1.465741 &0.917 &-1.376351 \\
0 & 3 &3  &3.0843 &1.097 &-1.144493 &1.020 &-1.192417 &0.919 & -1.062554 \\
\hline
-0.25 & 1.75 & 3.0857 & 8.8122 &1.100 &-6.003057 &1.000 &-5.784407 &0.903 &-6.154037\\
-0.25 & 2 &3.0857 &6.7468 &1.103 &-4.42299 &1.007 &-4.292183 &0.907 &-4.508668\\
-0.25 & 2.25 &3.0857 &5.3308 &1.093 &-3.3654 &1.012 & -3.302209 &0.911 &-3.400327\\
-0.25 & 2.5 &3.0857 &4.3180 &1.085 & {\bf -2.632778} &1.015 &-2.597435 &0.914 & -2.602943\\
-0.25 & 2.6 &3.0857 &3.9922 &1.091 & {\bf -2.39750} &1.016 &-2.376423 &0.916 &-2.351800\\
-0.25 & 2.75 &3.0857 &3.5686 &1.043 &-2.08964 &1.017 & -2.092873 &0.917 &-2.046106\\
-0.25 & 3 &3.0857 &2.9986 &1.096 &-1.672261  &1.019 &-1.714777 &0.919 &-1.614218\\
\hline
-0.5 & 1.75 & 3.1765 &8.5604 &1.026 & -5.101 (CHOL) &0.996 &-7.662615 &0.901 &-8.166828\\
-0.5 & 2 & 3.1765 &6.5540 &1.130 &-5.872775  &1.004 &-5.690254 &0.907 &-6.016180\\
-0.5 & 2.5 & 3.1765 &4.1946 &1.069 &  -3.501575 &1.013 &-3.472278 &0.913 &-3.546119\\
-0.5 & 2.75 & 3.1765 &3.4666 &1.050 & -2.770071 &1.016 &-2.812765 &0.915 &-2.798638\\
-0.5 & 3 & 3.1765 &2.9129 &1.045 & -2.250650 &1.018 &-2.319368 &0.918 &-2.2350\\
\hline
\end{tabular}
\caption{Final free energy densities $\tilde f(r) \equiv 10^5 f$ in LBU. These are fully equilibrated (with redshift $r$ as shown) for the ordered phases, and are redshifted also for the runs initialized from double twisted droplets in a cholesteric background for BPIII. Bold entries mark the free energies where BPIII is more stable than any of the ordered phases. (BPI is not competitive in this region of parameter space.) Note that for all parameters shown here BPIII is kinetically stable when initialized as described, with the exception of one low chirality run (marked CHOL) which reverted to a pure cholesteric state.}
\end{table*}

\subsection{Dielectric anisotropy effect}

In the main text, we have considered the response of our candidate blue phase III structure to an electric field, assuming that the medium had a positive dielectric constant. We have also considered the case of negative dielectric constant, which favors an ordering of the director field normal to the electric field direction.  Also in this case we observe ordering, although the final state is different, and consists of double helical disclinations parallel to the direction of the field. This state is shown in Fig.~6.

\begin{figure*}
\includegraphics[width=0.45\columnwidth]{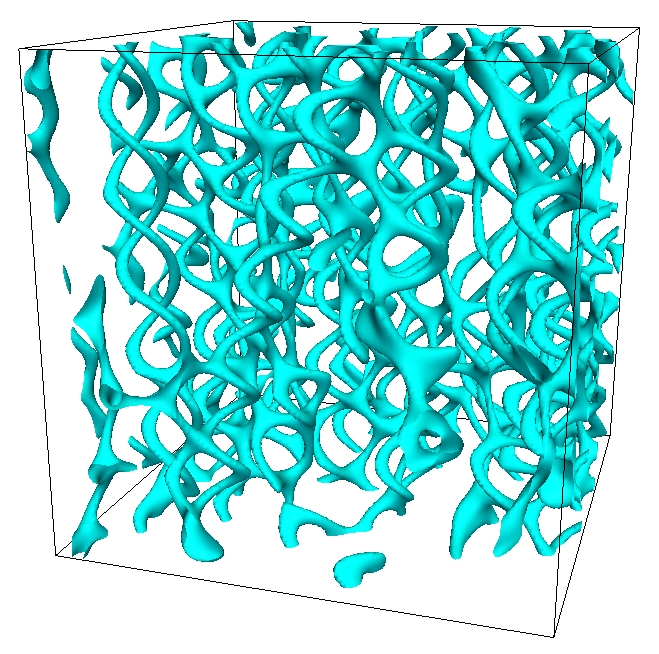}
\caption{Field-induced structure obtained when blue phase III is subjected to an electric field along the $z$ axis, in the regime analogous to the one explored in Fig.~3 of the main text, but this time with negative dielectric constant.}
\end{figure*}


\begin{thebibliography}{99}
\bibitem{supercomputers} Our runs were parallel, e.g. 10$^5$ LB time steps for a 128$^3$ lattice, on 512 nodes, required 3 hours (on the Hector supercomputer).
\end{thebibliography}
\end{document}